\documentclass[12pt,letterpaper]{article}

\usepackage[margin=1in]{geometry}
\usepackage{amsmath,amssymb}
\usepackage{booktabs}
\usepackage{multirow}
\usepackage[round,authoryear]{natbib}
\usepackage{indentfirst}
\usepackage{placeins}
\usepackage{setspace}
\usepackage{hyperref}
\usepackage{microtype}
\usepackage{parskip}
\usepackage{graphicx}

\hypersetup{
    colorlinks=true,
    linkcolor=blue,
    citecolor=blue,
    urlcolor=blue,
    pdftitle={Forecasting State Employment Relativity with Health, Labor Flow, and Small Establishment Indicators},
    pdfauthor={Dingyuan Liu and Yiqing Wang}
}

\begin{document}

\begin{titlepage}
\centering
\vspace*{1.5cm}

{\Large\bfseries A Counterfactual Diagnostic Framework for Explaining KS Deterioration in Credit Risk Model Validation}

\vspace{1.5cm}

{\large
Yiqing Wang\footnotemark\\[4pt]
\textit{Independent Researcher}\\[4pt]
Dallas, TX, USA
}

\footnotetext{Dr.Yiqing Wang is now a Model Validator at Citigroup. The views expressed in this paper are solely those of the author and do not reflect the views of Citigroup or any of its affiliates. Contact Email: woshilucy712@gmail.com}

\vspace{2cm}

\begin{abstract}
The Kolmogorov–Smirnov (KS) statistic is widely used in credit risk model monitoring and validation to assess discriminatory power. In practice, a material decline in KS often triggers governance review and requires validation teams to identify the breach source and the potential business risk. However, such diagnosis is frequently conducted on an ad hoc basis, relying on the judgment of individual validators rather than a standardized analytical framework. This paper proposes a counterfactual diagnostic framework for explaining KS deterioration in credit risk model validation. The framework sequentially attributes observed KS decline to sampling variability, portfolio composition change, covariate shift, and residual deterioration consistent with model drift, with explicit gateway conditions governing escalation at each stage. Simulation experiments demonstrate that the proposed approach provides more interpretable and governance-relevant explanations than threshold-based review alone, and contributes to more consistent, transparent, and defensible performance-breach assessment in credit risk model validation.
\end{abstract}

\vspace{1cm}

\noindent\textbf{Keywords:}  Model Risk Management; Credit Risk Model Validation; KS statistic

\vspace{0.5cm}

\end{titlepage}


\section{Introduction}
\label{sec:intro}

Credit scoring models are central to consumer lending decisions, and 
their ongoing performance is subject to continuous regulatory scrutiny. 
Under SR~11-7 \citep{SR117}, financial institutions are required to 
monitor deployed models against pre-specified performance benchmarks 
and to investigate any material deterioration in a structured and 
auditable manner. In credit risk, a breach of a governance 
threshold on the KS percentage change is a standard trigger for formal 
root-cause analysis. 

An observed KS deterioration may reflect several fundamentally distinct phenomena. Each carries fundamentally different governance implications, ranging from no action to model redevelopment. A diagnostic framework that conflates these drivers, or investigates them without a principled ordering, risks both under-reaction to genuine model failure and over-reaction to composition-driven noise. 

To address this risk, we propose a four-step sequential diagnostic framework that attributes observed KS deterioration to sampling variability, business composition change, covariate shift, or intrinsic model deterioration in a principle order. Each step produces a quantitative gateway condition that determines whether escalation is warranted, providing a structured, reproducible, and auditable basis for governance decisions. The framework makes two major contributions. First, it introduces a formal KS decomposition for different potential effects. Second, it provides a sequential gateway structure that maps directly onto SR~11-7 
governance requirements, making it immediately actionable within existing model risk management infrastructures.

The remainder of the paper is organized as follows. Section~\ref{sec:lit_review} reviews the relevant literature. Section~\ref{sec:method} presents the four-step diagnostic framework in detai. Section~\ref{sec:results} reports simulation results validating each diagnostic step. Section~\ref{sec:conclusion} concludes and discusses directions for future research.

\section{Literature Review}
\label{sec:lit_review}

Model risk management in financial institutions is governed by a 
well-established regulatory framework. The Federal Reserve and the 
Office of the Comptroller of the Currency jointly issued Supervisory 
Guidance SR~11-7 \citep{SR117}, which formalizes supervisory 
expectations for model development, validation, and ongoing monitoring. 
Under SR~11-7, effective validation requires not only an evaluation of 
conceptual soundness at the point of development, but also continuous 
monitoring of model performance over time to detect deterioration 
relative to design objectives. In particular, the guidance explicitly 
requires institutions to assess whether changes in products, exposures, 
or market conditions necessitate model adjustment, recalibration, or 
redevelopment. Despite this regulatory mandate, SR~11-7 provides 
principles-based guidance rather than prescriptive methodology, leaving 
financial institutions to develop their own operational frameworks for 
performance breach diagnosis. 

The Kolmogorov-Smirnov (KS) statistic is one of the most widely adopted 
discriminatory performance metrics in consumer credit risk modeling 
\citep{hand1997, thomas2002}. It is defined as the maximum absolute difference between the empirical cumulative distribution functions of model scores for goods and bads, and measures the extent 
to which a scoring model separates the two risk classes across the full 
score range. It  identifies the score threshold at which 
separation is maximized, making it particularly interpretable for 
cutoff strategy and approval policy design \citep{rezac2011}. In the 
context of ongoing performance monitoring, deviations of the KS 
statistic from a reference value are used to trigger 
governance review \citep{anderson2007}. However, the observed KS change 
between two monitoring periods reflects a mixture of effects, including 
changes in business composition, distributional shifts in model inputs, 
and intrinsic model deterioration. Disentangling these effects is not 
straightforward, and the existing credit scoring literature has not, to 
our knowledge, proposed a systematic decomposition framework for this 
purpose. 

A persistent practical challenge in credit risk model monitoring is 
distinguishing genuine model deterioration from apparent performance 
changes attributable to shifts in business composition. Regulatory 
The OCC \citep{occ1997} 
notes that observed deviations from reference performance may stem from 
changes in the lender's strategy rather than the model's true 
discriminatory power, and that these sources must be disentangled before 
management action is warranted. Similarly, \citet{thomas2002} and 
\citet{anderson2007} discuss the dependence of aggregate scorecard 
performance on the composition of the scored population, noting that 
cutoff adjustments and product or channel changes alter the mix of 
approved applicants in ways that mechanically affect observed KS  even when the model's intrinsic rank-ordering ability 
is unchanged. \citet{kritzinger2019} address this issue in the context of 
application scorecard monitoring, introducing a swap-set adjustment to 
the Gini coefficient that re-expresses post-implementation performance 
on a basis comparable to the development window, thereby removing the 
distortion caused by population substitution following scorecard 
implementation. Their approach demonstrates the conceptual importance 
of constructing composition-adjusted performance benchmarks, though it 
focuses on a specific type of policy change rather than providing a general decomposition framework 
applicable to arbitrary product or channel mix shifts. More broadly, 
the credit scoring literature has established that portfolio 
segmentation creates systematic heterogeneity in model performance 
across sub-populations \citep{thomas2002, bijak2012}. When the 
relative weights of these sub-populations change over time, the 
aggregate KS statistic reflects a mixture of within-segment and 
between-segment effects. 

A fundamental challenge in the deployment of statistical models is the 
divergence between the distribution of training data and the 
distribution of data encountered at the time of prediction, a 
phenomenon broadly referred to as dataset shift \citep{quinonero2009}. 
Within this general class, covariate shift refers specifically to the 
case where the marginal distribution of input features changes between 
the source and target domains while the conditional distribution of the 
outcome given inputs remains stable \citep{shimodaira2000}. This 
distinction is theoretically important: under covariate shift, a model 
that was well-specified at development may exhibit degraded predictive 
performance in deployment, not because the underlying relationship has 
changed, but because the population it encounters has moved to a region 
of the feature space that is underrepresented in the training data. The 
standard approach for addressing covariate shift is importance 
weighting, whereby training observations are reweighted by the ratio of 
target to source covariate densities, so that the reweighted source 
distribution matches the target \citep{shimodaira2000, sugiyama2007}. 
Direct estimation of this density ratio, without separately estimating 
the two densities, has been shown to improve robustness in 
high-dimensional settings \citep{sugiyama2008}. In practice, the 
density ratio is often approximated by a probabilistic classifier 
trained to distinguish source from target observations, with the 
log-odds of the classifier providing a consistent estimate of the log 
density ratio \citep{quinonero2009}. The relevance of covariate shift 
to credit risk model degradation has been recognized in both the 
academic and practitioner literature \citep{qian2022}, where 
distributional divergence between training and deployment populations 
has been identified as a key driver of scorecard performance decay. 

Finally, within the dataset shift literature, a conceptually important distinction separates covariate shift from concept drift, where the latter refers to changes in the conditional relationship between inputs and the outcome rather than changes in the input distribution alone \citep{gama2014}. The diagnostic logic of the proposed framework maps directly onto this distinction. Steps~2 and~3 sequentially test whether the observed KS breach can be explained by composition change and covariate shift respectively, while Step~4 attributes any remaining unexplained gap to intrinsic model degradation, consistent with the concept-drift interpretation. To our knowledge, no prior work has addressed this distinction within a sequential, gateway-structured diagnostic framework for credit risk model validation.

\section{Diagnosis Framework}
\label{sec:method}

\begin{figure}[htbp]
    \centering
    \includegraphics[width=0.8\textwidth,height=11cm, keepaspectratio]{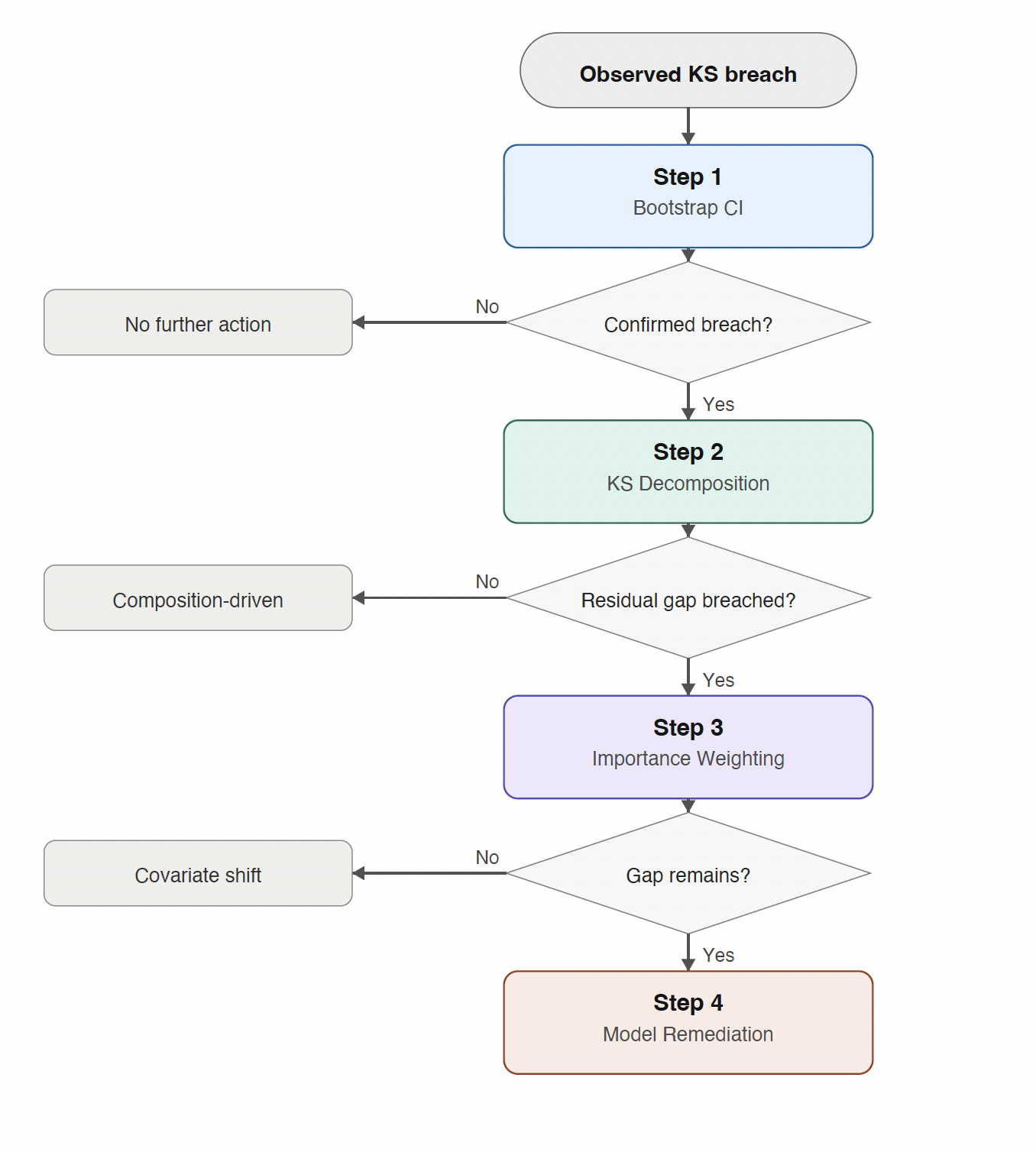}
    \caption{Diagnostic Framework}
    \label{fig:framework}
\end{figure}

This section presents the four-step sequential diagnostic framework for explaining KS deterioration in  credit risk model validation as summarized in Figure~\ref{fig:framework}. Each step produces a quantitative gateway condition that determines whether the observed breach can be explained at the current level of analysis or must be escalated to the next. 

\subsection{Step~1: Statistical Confirmation of Observed KS Deterioration}
\label{subsec:step1_significance}

Before conducting any root-cause diagnosis, we first determine whether the observed change in KS reflects a genuine deterioration rather than transient sampling variability. We denote $KS_{ref}$ and $KS_{cur}$
as the observed KS statistics in the reference and current periods, respectively. Following the common monitoring practice, we define the observed KS percentage change as
\begin{equation}
\label{eq:ks_pct_obs}
\%\Delta KS
=
\frac{KS_{cur}-KS_{ref}}{KS_{ref}}.
\end{equation}
Without loss of generality, and consistent with common monitoring conventions, we restrict our discussion to the case where $\%\Delta KS < 0$. The results could extend straightforwardly to the positive side.

To quantify sampling uncertainty, we employ stratified bootstrap resampling strategy separately within the reference and current samples. Specifically, good and bad observations are resampled with replacement while preserving the original class counts in each period. For each bootstrap replication $b=1,2,\dots,B$, we recompute $KS_{ref}^{(b)}$, $KS_{cur}^{(b)}$
and the corresponding bootstrap KS percentage change
\begin{equation}
\label{eq:ks_pct_bootstrap}
\%\Delta KS^{(b)}
=
\frac{KS_{cur}^{(b)}-KS_{ref}^{(b)}}{KS_{ref}^{(b)}}.
\end{equation}

The empirical distribution of $\{\%\Delta KS^{(b)}\}_{b=1}^B$ is then used to construct a bootstrap confidence interval at level $1-\alpha$:
\begin{equation}
\label{eq:ci_bootstrap}
CI_{1-\alpha}\!\left(\%\Delta KS\right)
=
\left[L_{\alpha/2},\, U_{1-\alpha/2}\right].
\end{equation}

This interval serves two purposes. First, it distinguishes statistically significant deterioration from ordinary sampling fluctuation. If $0$ is included in the interval, there is no sufficient evidence to reject the assumption that the KS percentage change is zero. Otherwise, the deterioration is deemed statistically significant. Second, the confidence interval can be compared against a pre-defined governance breach threshold $\tau$. A material breach is confirmed only if the entire confidence interval lies below $\tau$. If the interval lies below $0$ but overlaps $\tau$, then the deterioration is statistically significant, but the material breach is not yet confirmed. Table~\ref{table:step1_cases} summarize the decision logic for Step~1 with $\alpha=0.05$ and a pre-defined negative threshold $\tau$. 
\begin{table}[htbp]
\centering
\caption{Decision Scenarios for Step 1}
\label{table:step1_cases}
\resizebox{1\textwidth}{!}{
\begin{tabular}{|l|l|l|}
\hline
\multicolumn{1}{|c|}{\textbf{\shortstack{Case for \\ $CI_{95\%}(\%\Delta KS)=[L_{2.5\%},U_{97.5\%}]$}}}
& \multicolumn{1}{c|}{\textbf{Classification}}
& \multicolumn{1}{c|}{\textbf{Action}} \\ \hline
$\tau < L_{2.5\%} < 0 < U_{97.5\%}$
& No statistically supported deterioration
& No further analysis \\ \hline
$\tau < L_{2.5\%} < U_{97.5\%} < 0$
& Significant deterioration, no breach
& No further analysis \\ \hline
$L_{2.5\%} < \tau < U_{97.5\%} < 0$
& Significant deterioration, breach not confirmed
& Increasing monitoring frequency \\ \hline
$L_{2.5\%} < U_{97.5\%} < \tau$
& Confirmed material breach
& Proceed to root-cause analysis \\ \hline
\end{tabular}
}
\end{table}

Step 1 is to distinguish the material breach from observed breach. Once a material breach is confirmed, we could proceed to next steps to decompose the observed KS deterioration into different effects.

\subsection{Step~2: Policy-Driven Regime Changes}
\label{subsec:step2_policy_change}

Once the observed KS deterioration is both statistically significant and materially breached, Step~2 examines whether the deterioration can be explained by changes in business composition rather than by a decline in the model's intrinsic discriminatory power. In particular, the business regime in the current period may differ from that in the reference period because of product or channel entry and exit, as well as changes in the relative proportions of existing products or channels. The key objective of this step is therefore to isolate the  business structure effect from the overall KS breach.
We can decompose the observed the performance deterioration $KS$ into four components:
\begin{itemize}
    \item Current-only universe effect,
    \item Reference-only universe effect,
    \item Mix effect within common support
    \item Residual aligned performance gap.
\end{itemize}

\paragraph{Current-only universe effect.}
The current-only universe effect captures the performance of segments that appear in the current period but are absent from the reference period. These newly introduced products, channels, or regime combinations have no direct benchmark counterpart and therefore cannot be assessed through a apple-to-apple comparison. Their contribution should be interpreted as a business-regime expansion effect rather than direct evidence of intrinsic model deterioration.

\paragraph{Reference-only universe effect.}
The reference-only universe effect captures the impact of segments that were present in the reference period but are absent from the current period. Since these segments no longer belong to the current business structure, their historical contribution to reference KS should be excluded from the evaluation of current-period model performance.

\paragraph{Common-support product-mix-adjusted KS.}
After removing the current-only and reference-only universe effects, the remaining comparison is conducted on the common support, where differences in segment proportions can be further adjusted through mix alignment.To isolate the effect of changes in product composition, we first restrict attention to the set of products that are present in both the reference and current periods. 

Denote  $\mathcal G_B $ and $\mathcal G_C$ as the sets of observed segments in the reference and current periods, respectively. Their common support is defined as $\mathcal G^{com} = \mathcal G_B \cap \mathcal G_C$. For each segment $g \in \mathcal{G}_{com}$, let $N_g^{ref}$ and $N_g^{cur}$ denote the sample counts belonging to that segment during the reference and current periods, respectively. The product shares within the common support can be witted as
\[
\pi_g^{ref,com}
=
\frac{N_g^{ref}}{\sum_{h \in \mathcal{G}_{com}} N_h^{ref}},
\qquad
\pi_g^{cur,com}
=
\frac{N_g^{cur}}{\sum_{h \in \mathcal{G}_{com}} N_h^{cur}}.
\]

To align the reference sample with the current-period product composition, each reference observation belonging to product $g$ is assigned the weight
\[
w_g^{prod}
=
\frac{\pi_g^{cur,com}}{\pi_g^{ref,com}},
\qquad g \in \mathcal{G}_{com}.
\]

Let $s_i$ denote the model score for observation $i$, let $y_i \in \{0,1\}$ denote the class label, where $y_i=1$ indicates a bad and $y_i=0$ indicates a good, and let $g(i)$ denote the product membership of observation $i$. The weight for observation $i$ is then $w_i = w_{g(i)}^{prod}$. Using these weights, we construct the weighted empirical distribution functions for goods and bads in the reference sample:
\[
\hat{F}_0^{\,w}(t)
=
\frac{\sum_{i: y_i=0} w_i \mathbf{1}(s_i \le t)}
{\sum_{i: y_i=0} w_i},
\qquad
\hat{F}_1^{\,w}(t)
=
\frac{\sum_{i: y_i=1} w_i \mathbf{1}(s_i \le t)}
{\sum_{i: y_i=1} w_i}.
\]
The common-support product-mix-adjusted KS in reference sample is then defined as
\[
KS_{ref \to cur}^{com}
=
\sup_t
\left|
\hat{F}_1^{\,w}(t) - \hat{F}_0^{\,w}(t)
\right|.
\]

This quantity represents the KS that would have been observed in the reference period if the reference sample had the same product composition as the current sample within the common-support products. It therefore removes the component of the aggregate KS change that is attributable purely to shifts in product mix among products shared by both periods.

By considering the above effects, we can decompose the original observed KS difference into four components. Let  $KS_{ref}^{com}$ and $KS_{cur}^{com}$ denote the reference and current KS restricted to the common-support segments without mix distribution adjustment.
Then the observed KS change can be written as

\begin{align}
KS_{cur}-KS_{ref}
&=
\underbrace{\left(KS_{cur}-KS_{cur}^{com}\right)}_{\text{current-only universe effect}}
+
\underbrace{\left(KS_{cur}^{com}-KS_{ref \to cur}^{com,mix}\right)}_{\text{residual aligned performance gap}} \notag \\
&\quad+
\underbrace{\left(KS_{ref \to cur}^{com,mix}-KS_{ref}^{com}\right)}_{\text{mix effect within common support}}
+
\underbrace{\left(KS_{ref}^{com}-KS_{ref}\right)}_{\text{reference-only universe effect}}.
\end{align}

Equivalently, the originally observed percentage KS change can be decomposed using the same four components under a common denominator:
\[
\%\Delta KS
=
\frac{KS_{cur}-KS_{ref}}{KS_{ref}} \times 100\%
=
\%\Delta KS_{cur-only}
+
\%\Delta KS_{residual}
+
\%\Delta KS_{mix}
+
\%\Delta KS_{ref-only},
\]
where each term is defined by dividing the corresponding KS component above by $KS_{ref}$.

\paragraph{Interpretation and action for each component.}
The decomposition is not only explanatory, but also prescriptive in terms of subsequent governance actions. For each component, model validators and policy team should make different appropriate response.

For the \emph{current-only universe effect}, the primary implication is that the current portfolio includes newly introduced segments that have no direct reference counterpart. This component should therefore be managed as a regime-expansion effect rather than immediate evidence of intrinsic model deterioration. In practice, such segments should be monitored separately, and, where material, may require segment-specific cutoffs, policy overlays, temporary volume controls, or eventually a dedicated benchmark or separate modeling treatment if they become a stable part of the portfolio.

For the \emph{reference-only universe effect}, the implication is majorly related to reference comparability. Since these segments are no longer part of the current business structure, their historical contribution should be excluded from the evaluation of current-period model performance. The corresponding action is therefore to refine the reference comparison framework rather than to trigger model remediation.

For the \emph{mix effect within the common support}, the implication is that changes in the relative proportions of overlapping segments are materially affecting aggregate KS. In this case, governance focus should shift toward segment-level monitoring and portfolio steering, including possible revisions to segment-specific cutoffs, approval strategy, channel strategy, or ongoing monitoring benchmarks. If such mix shifts are persistent, then a mix-adjusted benchmark should be incorporated into regular monitoring.

For the \emph{residual aligned performance gap}, the implication is that a material deterioration remains even after removing universe effects and aligning segment composition. This should the only component that is carried forward as unexplained deterioration. Accordingly, it serves as the escalation trigger for next step investigation, where the remaining gap is further examined for covariate shift under the aligned business regime.

After removing universe effects and adjusting for mix change within the common support, we define the aligned residual percentage change as
\begin{equation}
\label{eq:aligned_residual_change}
\%\Delta KS_{aligned}
=
\frac{KS_{cur}^{com}-KS_{ref \to cur}^{com,mix}}
{KS_{ref \to cur}^{com,mix}} \times 100\%.
\end{equation}
This quantity measures the remaining deterioration after alignment to the current business structure. If $\%\Delta KS_{aligned}$ no longer breaches the governance threshold, then the originally observed KS deterioration is considered to be largely explained by business-composition change, and no further root-cause escalation is required under this framework. Otherwise, Step~3 should be triggered to investigate whether the remaining deterioration can be attributed to covariate shift rather than intrinsic model failure.

\subsection{Step~3: Covariate-shift Driven Change}
If the aligned residual percentage change (formula \ref{eq:aligned_residual_change}) from Step~2 remains materially breached, we further investigate whether the remaining deterioration can be explained by covariate shift, rather than by intrinsic model deterioration. In other words, we want to answer the question: \emph{if the reference sample were exposed to the current-period covariate distribution, what KS would it have achieved?} 

To answer this question, we reweight reference observations such that their covariate distribution matches that of the current period. We then recompute the reference KS under this covariate-aligned distribution. If the updated reference KS moves close to the observed current-period KS, the residual deterioration from Step~2 can be largely attributed to covariate shift, meaning that the current population has become harder to separate even though the model itself may not have intrinsically deteriorated. Otherwise, the covariate shift alone is insufficient to explain the deterioration, and further diagnosis is required in next steps.

At this stage, the comparison is restricted to the common-support segments, and the reference sample has already been aligned to the current segment composition through the Step~2 mix adjustment. The objective of Step~3 is therefore to assess whether the residual gap
\[
KS_{cur}^{com} - KS_{ref \to cur}^{com,mix}
\]
can be explained by differences in the covariate distribution within the aligned business regime.

Let $X=(X_1,\dots,X_p)$ denote the vector of model input variables that are consistently defined in both the reference and current periods. We estimate a covariate-shift weight by constructing a  domain classifier $f(.)$ on the common support $\mathcal G^{com}$, where observations from the current period are labeled by $Z=1$ and observations from the mix-adjusted reference sample are labeled by $Z=0$. The choice and tuning of the domain classifier should reflect the structure of the dataset and the complexity of the covariate 
shift being detected. Regardless of the classifier chosen, the area 
under the receiver operating characteristic curve (AUROC) provides a 
natural diagnostic for the degree of distributional divergence between the reference and current samples. An AUROC close to 0.5 indicates negligible covariate shift, while values substantially above 0.5 confirm that the two periods are distributionally distinguishable.

Let $\hat{p}(Z=1 \mid X_i, g_i)$
denote the fitted probability from the classifier $f(.)$, where $g_i$ is the specific segment of observation $i$ within the common support. Let $\eta = P(Z=1)$
denote the class proportion prior in the domain-classification sample. The covariate-shift weight is then defined as
\begin{equation}
\label{eq:cov_weight}
w_X(X_i,g_i)
=
\frac{1-\eta}{\eta}
\cdot
\frac{\hat{p}(Z=1 \mid X_i,g_i)}
{1-\hat{p}(Z=1 \mid X_i,g_i)}.
\end{equation}

For each reference observation $i$ in the common support, the combined weight from Step~2 and Step~3 is defined as
\[
w_i^{total}
=
w_{g(i)}^{prod}\cdot w_X(X_i,d_i),
\]
where $w_{g(i)}^{prod}$ is the Step~2 mixed weight. Using the overall weights, we construct the weighted empirical distribution functions for goods and bads in the reference sample:
\[
\hat{F}_0^{\,total}(t)
=
\frac{\sum_{i:y_i=0} w_i^{total}\mathbf{1}(s_i\le t)}
{\sum_{i:y_i=0} w_i^{total}},
\qquad
\hat{F}_1^{\,total}(t)
=
\frac{\sum_{i:y_i=1} w_i^{total}\mathbf{1}(s_i\le t)}
{\sum_{i:y_i=1} w_i^{total}},
\]
where $s_i$ denotes the model score and $y_i\in\{0,1\}$ denotes the performance label. The covariate-aligned benchmark KS is then defined as
\[
KS_{ref \to cur}^{com,mix,x}
=
\sup_t
\left|
\hat{F}_1^{\,total}(t)-\hat{F}_0^{\,total}(t)
\right|.
\]

This quantity represents the reference-period KS that would have been observed if the reference sample had both the current segment composition and the current covariate distribution within the common-support regime. We then define the remaining percentage change after covariate alignment as
\begin{equation}
\label{eq:KS_step3}
    \%\Delta KS_{x\text{-aligned}}
=
\frac{KS_{cur}^{com}-KS_{ref \to cur}^{com,mix,x}}
{KS_{ref \to cur}^{com,mix,x}}
\times 100\%.
\end{equation}
If $\%\Delta KS_{x\text{-aligned}}$ no longer breaches the governance threshold, then the remaining deterioration from Step~2 is considered largely explained by covariate shift. Otherwise, a material unexplained gap remains after both regime alignment and covariate alignment, and the diagnosis proceeds to Step~4.

\subsection{Step~4: Residual model-related deterioration.}
If a material breach remains after both regime alignment in Step~2 and covariate alignment in Step~3, the remaining deterioration is treated as unexplained by business-composition change or covariate shift. 

We then consider the residual deterioration as evidence of model-related degradation, which may arise from several potential sources. First, the relationship between attributes and response variable may have changed over time, and hence patterns learned in the reference period no longer hold in the current period. Second, the model's score ordering may have weakened, leading to reduced separation between goods and bads even under an aligned population structure. Third, the model specification itself may be insufficient for the current environment, for example because important nonlinearities, interactions, or newly relevant risk patterns are not adequately captured.

Accordingly, the output of Step~4 is not a further decomposition of KS change, but a governance review and remediation. Depending on severity and business context, appropriate actions may include recalibration, challenging model analysis, feature review, segmentation redesign, or full model redevelopment. In this sense, Step~4 serves as the final escalation stage of the diagnostic framework. After removing regime-driven and population-driven explanations, any remaining material KS deterioration is treated as evidence that the model itself no longer performs adequately under the current environment.

\section{Simulation Results}
\label{sec:results}

In this section, we evaluate the proposed framework through controlled simulation experiments from Steps~1 through~3. For each step, synthetic datasets are constructed to isolate distinct deterioration drivers, allowing the diagnostic logic to be assessed under known ground truth. The results demonstrate that the sequential procedure reliably distinguishes among sampling variability, policy-driven regime change, and covariate shift, correctly halting or escalating the diagnosis at each gateway in accordance with the underlying data-generating process.

\subsection{Step~1 Breach Confirmation}
We illustrate Step 1 using four synthetic scenarios designed to span the complete decision space defined in Table \ref{table:step1_cases}. In each scenario, reference and current samples are drawn from a logistic score-generating process with controlled separation parameters, and 1,000 stratified bootstrap replications are used to construct a $95\%$ confidence interval for $\%\Delta KS$. The governance breach threshold is set at $\tau = -20\%$. 

\begin{figure}[htbp]
    \centering
    \includegraphics[width=0.8\textwidth]{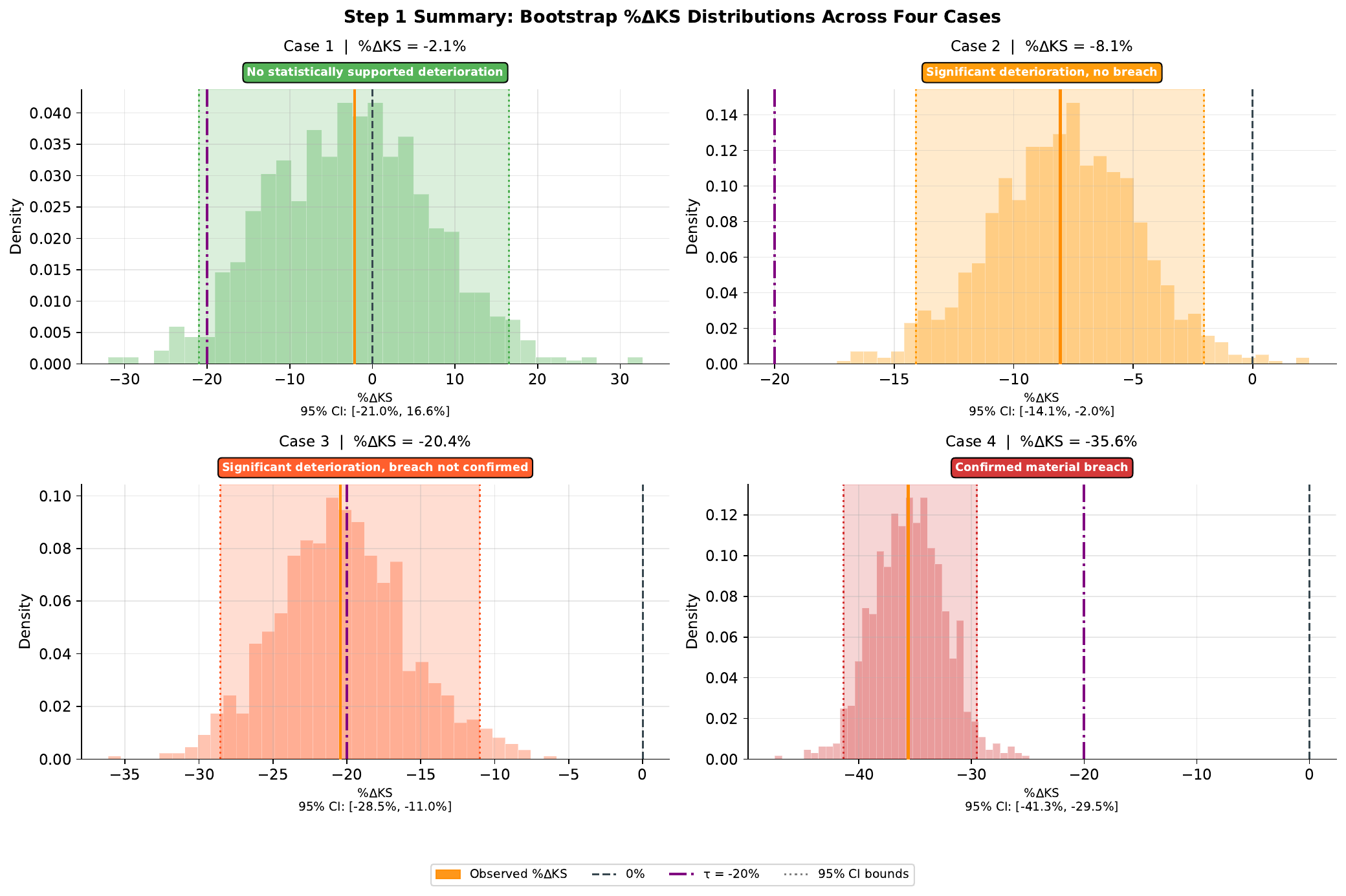}
    \caption{Simulation Result for Step 1}
    \label{fig:step1}
\end{figure}

Figure~\ref{fig:step1} illustrates the four governance outcomes using simulated examples. Case~1 produces a wide confidence interval across value zero, yielding no statistically supported deterioration despite a negative point estimate. Case~2 shows a statistically significant decline of KS, but the interval remains above $\tau=-20\%$, so no governance breach is triggered. Case~3 has a more severe drop of $\%\Delta KS=-20.5\%$ whose confidence interval straddles $\tau$, yielding a breach-not-confirmed classification that calls for increased monitoring rather than immediate remediation. Case~4 represents a confirmed material breach, with the entire interval falling well below $\tau$. Taken together, the four cases demonstrate that statistical significance and material breach are distinct criteria whose interplay depends on both the magnitude of deterioration and the precision of the bootstrap estimate.

\subsection{Step~2 Regime Change Decomposition}

To validate the regime change decomposition, we construct four synthetic datasets, where each is engineered to activate a distinct subset of the decomposition terms. All datasets are generated from a latent logistic model in which the KS statistic is governed by a separation parameter (Sep) controlling the mean difference between the good and bad score distributions. Within-segment KS is monotonically increasing in separation, while segment-level feature have no effect on KS.

\begin{table}[htbp]
\centering
\caption{Synthetic dataset configurations for Step 2 decomposition validation}
\label{tab:step2_scenarios}
\renewcommand{\arraystretch}{1.25}
\begin{tabular}{llcccc}
\hline
\textbf{Scenario} & \textbf{Segment} & \textbf{Ref \%} & \textbf{Cur \%} &
  \textbf{Ref Sep.*} & \textbf{Cur Sep.} \\
\hline
\multirow{2}{*}{S2-A: Pure Mix Shift}
  & A  & 70 & 30 & 2.5 & 2.5 \\
  & B   & 30 & 70 & 1.0 & 1.0 \\
\hline
\multirow{3}{*}{S2-B: Universe Change}
  & C (exits)  & 40 &  — & 2.5 &  — \\
  & D (enters) &  — & 30 &  — & 1.2 \\
  & E (common) & 60 & 70 & 2.0 & 2.0 \\
\hline
\multirow{2}{*}{S2-C: Pure Residual Gap}
  & A & 50 & 50 & 2.5 & 1.2 \\
  & B & 50 & 50 & 2.0 & 0.9 \\
\hline
\multirow{4}{*}{S2-D: Mixed Effects}
  & A (common)  & 50 & 30 & 2.5 & 2.0 \\
  & B (common)  & 30 & 40 & 2.0 & 1.8 \\
  & C (exits)   & 20 &  — & 1.5 &  — \\
  & D (enters)  &  — & 30 &  — & 1.2 \\
\hline
\multicolumn{6}{l}{\footnotesize $^*$ ``—'' denotes that the segment is absent in that period.} \\
\multicolumn{6}{l}{\footnotesize $^*$ Sep*. governs within-segment discriminatory power (higher Sep means better KS).} \\

\end{tabular}
\end{table}

\begin{figure}[htbp]
    \centering
    \includegraphics[width=0.8\textwidth]{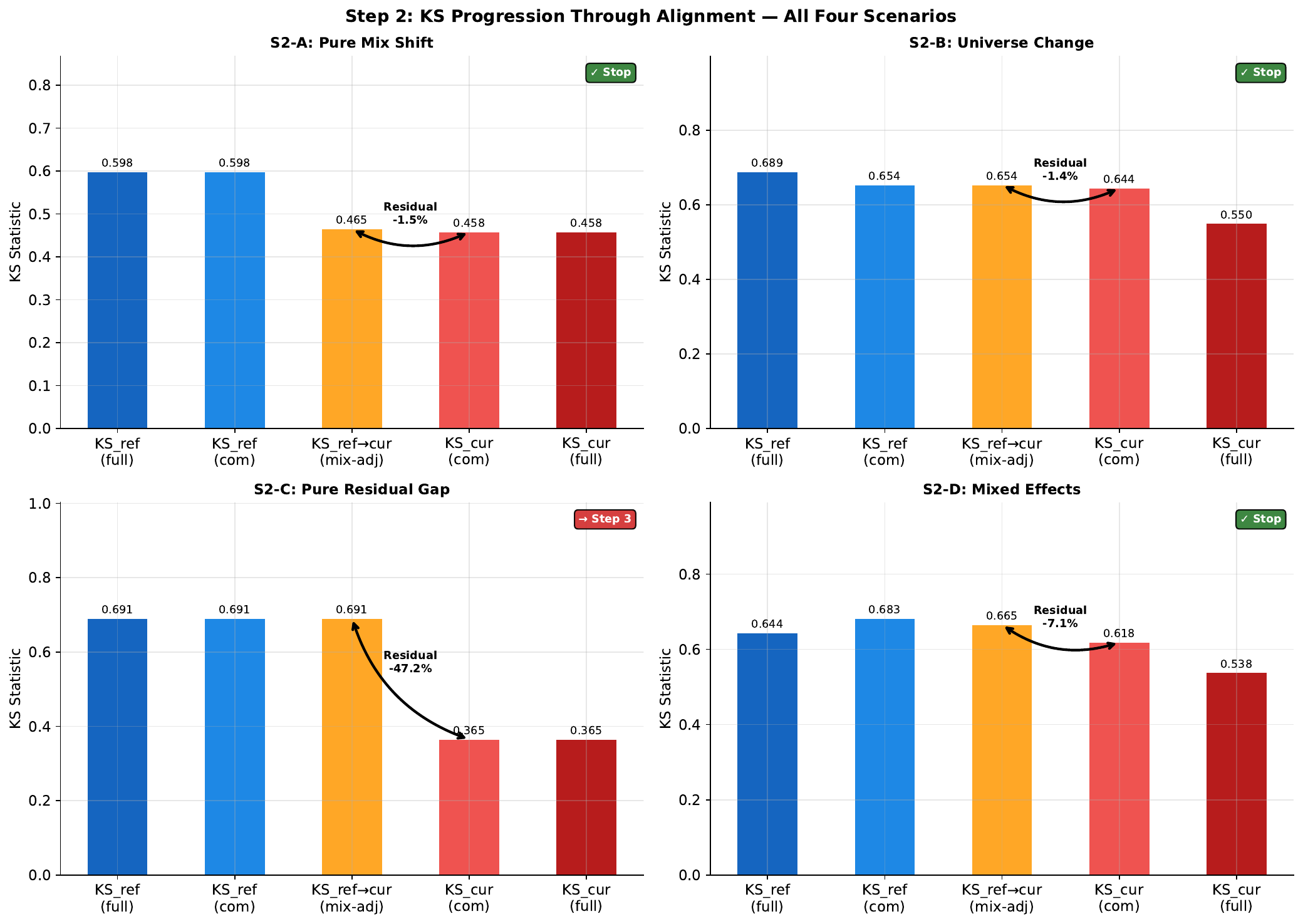}
    \caption{Simulation Result for Step 2}
    \label{fig:step2}
\end{figure}

Figure~\ref{fig:step2} illustrates the Step~2 decomposition across four simulated scenarios corresponding the four components of KS change. 
\begin{itemize}
    \item In S2-A pure mix shift, a reversal of segment proportions (70:30 to 30:70 between two segments) produces a significant KS decline from $59.8\%$ to $45.8\%$, driven almost entirely by the product mix effect. After reweighting, the residual aligned gap change is only $-1.5\%$ and the pipeline correctly halts at Step~2. 
    \item In S2-B universe change, The change of segment types account for the KS decline from $68.9\%$ to $55.0\%$. The residual aligned gap change is $-1.4\%$ and hence the pipeline halts at Step 2. As the KS decreases is majorly contributed by new product entry in current sample and original product exit, we need to evaluate those specific products' performance separately. 
    \item In S2-C pure residual gap, both segments remain with identical proportions but within-segment separation collapses, producing a residual aligned gap of $47.2\%$ with  all other structural components zero. As the KS change is triggering the threshold, the pipeline escalates to Step~3. 
    \item In S2-D Mixed Effects, we combines all drivers simultaneously: universe effects, a small mix shift, and within-segment deterioration together produce an aggregate decline of $-16.5\%$, of which the residual aligned gap accounts for $-7.1\%$
\end{itemize}

Collectively, the four simulation scenarios demonstrate that the Step~2 decomposition reliably isolates business-composition effects from residual model-related deterioration, correctly attributing KS changes to their respective drivers in each case.

\subsection{Step~3 Covariate Shift Detection}

\begin{figure}[htbp]
    \centering
    \includegraphics[width=0.8\textwidth]{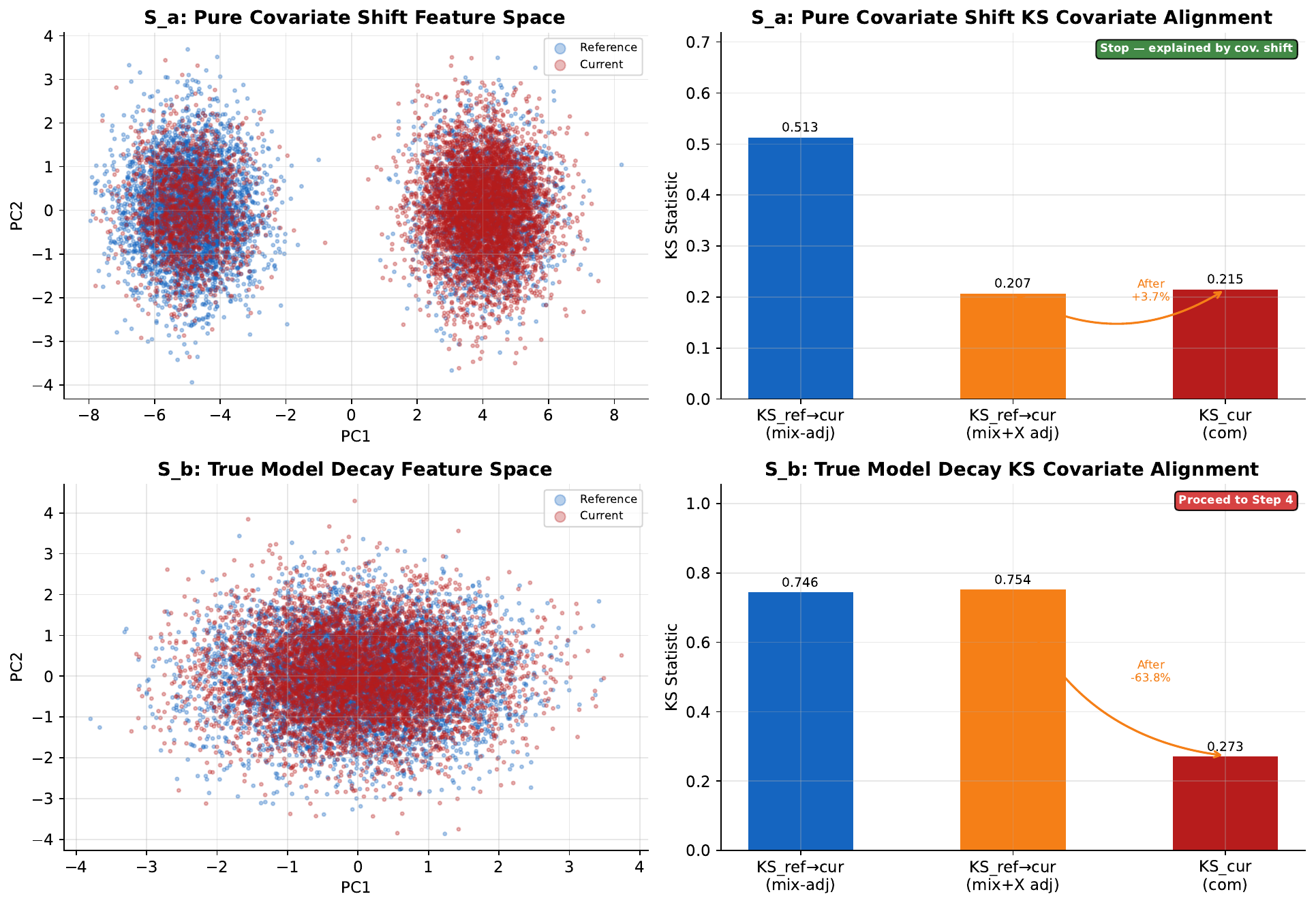}
    \caption{Simulation Result for Step 3}
    \label{fig:step3}
\end{figure}

Figure~\ref{fig:step3} and Table ~\ref{tab:step3_scenarios} summarizes the covariate alignment check for two simulated scenarios. In scenario a, we make the share of a high risk sub-population rises from $35\%$ to $75\%$ from reference to current periods, driving an aggregate KS decline from $51.3\%$ to $21.5\%$ with no 
change in the model itself. The domain classifier yields an AUC of 
0.772, showing a detectable covariate shift. After importance 
weighting, the reweighted reference KS is $20.7\%$, which is aligned with the current KS, and we can claim the KS decrease is majorly due to the covariate shift.
In another scenario b, the reference and current feature distributions are identical by construction, but within-sample separation decreases, simulating genuine model decay. The domain classifier has an AUC of 0.644, close to the uninformative baseline, and the adjusted KS keep unchanged from $74.6\%$ to $75.4\%$, leading to the next step detection. Together, the two scenarios illustrate that Step~3 differentiate cleanly between covariate-driven and model-driven deterioration.

\begin{table}[ht]
\centering
\caption{Step 3 Simulation Results}
\label{tab:step3_scenarios}
\begin{tabular}{|l|llll|ll|}
\hline
\multirow{2}{*}{\textbf{Scenarios}} & \multicolumn{4}{l|}{\textbf{Step 2: Domain Diagnosis}}                                                                                                                                                     & \multicolumn{2}{l|}{\textbf{Step 3: Covariate Diagnosis}}                                                                    \\ \cline{2-7} 
                           & \multicolumn{1}{l|}{$KS_{ref}$} & \multicolumn{1}{l|}{$KS_{cur}$} & \multicolumn{1}{l|}{$KS_{ref \to cur}^{com,mix}$} & $KS_{cur}^{com}$ & \multicolumn{1}{l|}{AUROC of domain classifier} & $KS_{ref \to cur}^{com,mix,x}$ \\ \hline
a                          & \multicolumn{1}{l|}{51.3\%}      & \multicolumn{1}{l|}{21.5\%}      & \multicolumn{1}{l|}{51.3\%}                                                          & 21.5\%                               & \multicolumn{1}{l|}{0.772}                     & 20.7\%                                                            \\ \hline
b                          & \multicolumn{1}{l|}{74.6\%}      & \multicolumn{1}{l|}{27.3\%}      & \multicolumn{1}{l|}{74.6\%}                                                          & 27.3\%                               & \multicolumn{1}{l|}{0.644}                     & 75.4\%                                                            \\ \hline
\end{tabular}
\end{table}

\section{Conclusions}
\label{sec:conclusion}

We propose a four-step sequential diagnostic framework for explaining KS deterioration in credit risk model validation. By combining bootstrap-based uncertainty assessment, regime-level KS decomposition, 
and covariate weighting, the framework provides a structured and reproducible basis for performing root cause analysis of KS considering different components. Each step produces a quantitative gateway condition 
that governs escalation, ensuring that governance action is proportionate to the identified root cause. Simulation experiments illustrate that the framework reliably distinguishes among these drivers under controlled 
conditions, correctly halting or escalating the diagnosis at each stage.

The work contributes to model risk management practice in two major respects. Methodologically, it introduces a formal KS decomposition that separate mix effect and residual effect, and introducing importance weighting within the specific objective of counterfactual KS estimation. Practically, the sequential gateway structure maps directly onto SR~11-7 governance requirements, making the framework immediately actionable within existing model risk management infrastructures and providing auditable documentation for regulatory review.

Several directions remain open for future work. First, the current framework addresses only the case of KS deterioration. Extending the analysis to settings where KS increases significantly would broaden its applicability. Second, the choice of domain classifier in Step~3 is left to the practitioner. A systematic comparison of classifier families under varying shift intensities and sample sizes would provide more prescriptive guidance for users. Finally, the framework is developed and evaluated in the context of credit risk model validation. Assessing its applicability to other model types and risk domains would further establish its robustness and governance relevance in broader real-world deployments.

\section*{Statements and Declarations}

\medskip
\noindent\textbf{Funding.} This research received no external funding.

\medskip
\noindent\textbf{Competing interests.} The authors declare no competing interests.

\medskip
\noindent\textbf{AI usage.} The authors used Claude Sonnet 4.6 (Anthropic) for assistance with manuscript polishing only. 

\bibliographystyle{plainnat}

\bibliography{references}

\end{document}